\documentclass[superscriptaddress,prb,nobalancelastpage,tightenlines,nofootinbib]{revtex4-2}
\usepackage{amssymb}
\usepackage{amsmath}
\usepackage{array}
\usepackage{graphicx}
\usepackage{tikz}
\usepackage{pgfplots}
\usepackage{placeins}
\usepackage{relsize}
\usepackage{tabu,multirow}
\usepackage{epsfig}
\usepackage{latexsym}
\usepackage{subfigure}
\usepackage{soul}
\usepackage[export]{adjustbox}
\usepackage{setspace}
\usepackage{float}
\usepackage{stmaryrd}
\usepackage{miller}
\usepackage{bm}
\usepackage{tabularx}
\usepackage{booktabs}
\usepackage[hidelinks]{hyperref} 
\usepackage[noabbrev,capitalize]{cleveref}
\usepackage{ragged2e}
\usepackage{titlesec}
\usepackage{multirow}
\usepackage{makecell}
\graphicspath{{./}}

\begin{document}
\title{Refining interface stress measurement in nanomultilayers through layer corrugation and interface roughness corrections}
\author{Yang Hu}
\affiliation{Laboratory for Advanced Materials Processing, Empa - Swiss Federal Laboratories for Materials Science and Technology, Thun, Switzerland\\}
\author{Aleksandr Druzhinin}
\affiliation{Laboratory for Joining Technologies and Corrosion, Empa - Swiss Federal Laboratories for Materials Science and Technology, Duebendorf, Switzerland\\}%
\author{Claudia Cancellieri*}
\affiliation{Laboratory for Joining Technologies and Corrosion, Empa - Swiss Federal Laboratories for Materials Science and Technology, Duebendorf, Switzerland\\}%
\author{Vladyslav Turlo*}
\affiliation{Laboratory for Advanced Materials Processing, Empa - Swiss Federal Laboratories for Materials Science and Technology, Thun, Switzerland\\}%
\affiliation{National Centre for Computational Design and Discovery of Novel Materials MARVEL, Empa, Thun, Switzerland\\}%
\email{claudia.cancellieri@empa.ch, vladyslav.turlo@empa.ch}

\date{\today}

\begin{abstract}
We introduce new models that incorporate layer corrugation and interface roughness into standard approaches for measuring interface stress in nanomultilayers (NMLs). Applied to Cu/W NMLs, these models show that ignoring such features can inflate measured interface stress by up to 0.4 J/m$^2$. However, corrugation and roughness alone cannot account for the extreme stresses reported, suggesting that atomic-scale phenomena (e.g., intermixing and metastable phase formation at the interfaces) dominate. These findings highlight the importance of balancing bilayer counts and thickness-to-roughness ratios for reliable stress quantification, providing a practical pathway to designing and characterizing advanced nanocomposite coatings with improved accuracy.
\end{abstract}


\maketitle

\section{Introduction}

Nano-multilayers (NMLs) are nanocomposite coatings composed of two or more distinct phases. They exhibit a high density of heterogeneous interfaces, which imparts unique magnetic, optical, mechanical, and radiation-tolerant properties. Consequently, NMLs find broad application in microelectronics \cite{auciello1991controlled,kim2003multilayer,nam2001improved,ono1994diffusion}, micro-/nanojoining \cite{janczak2014interfacial,kaptay2016melting,baras2016,baras2018shs}, optical coatings \cite{anders1993vacuum,stueber2009concepts}, radiation-tolerant coatings \cite{gao2011radiation}, and micro-/nanoelectromechanical systems \cite{Trevizo2020}, among others. Meanwhile, the intrinsic stress developed during NML growth plays a critical role in determining thermal stability, functional behavior, and mechanical performance \cite{lorenzin2022tensile,lorenzin2024effect}.

\begin{figure}[H]
\centering
\includegraphics[width=0.75\textwidth,clip,trim=0cm 0cm 0cm 0cm]{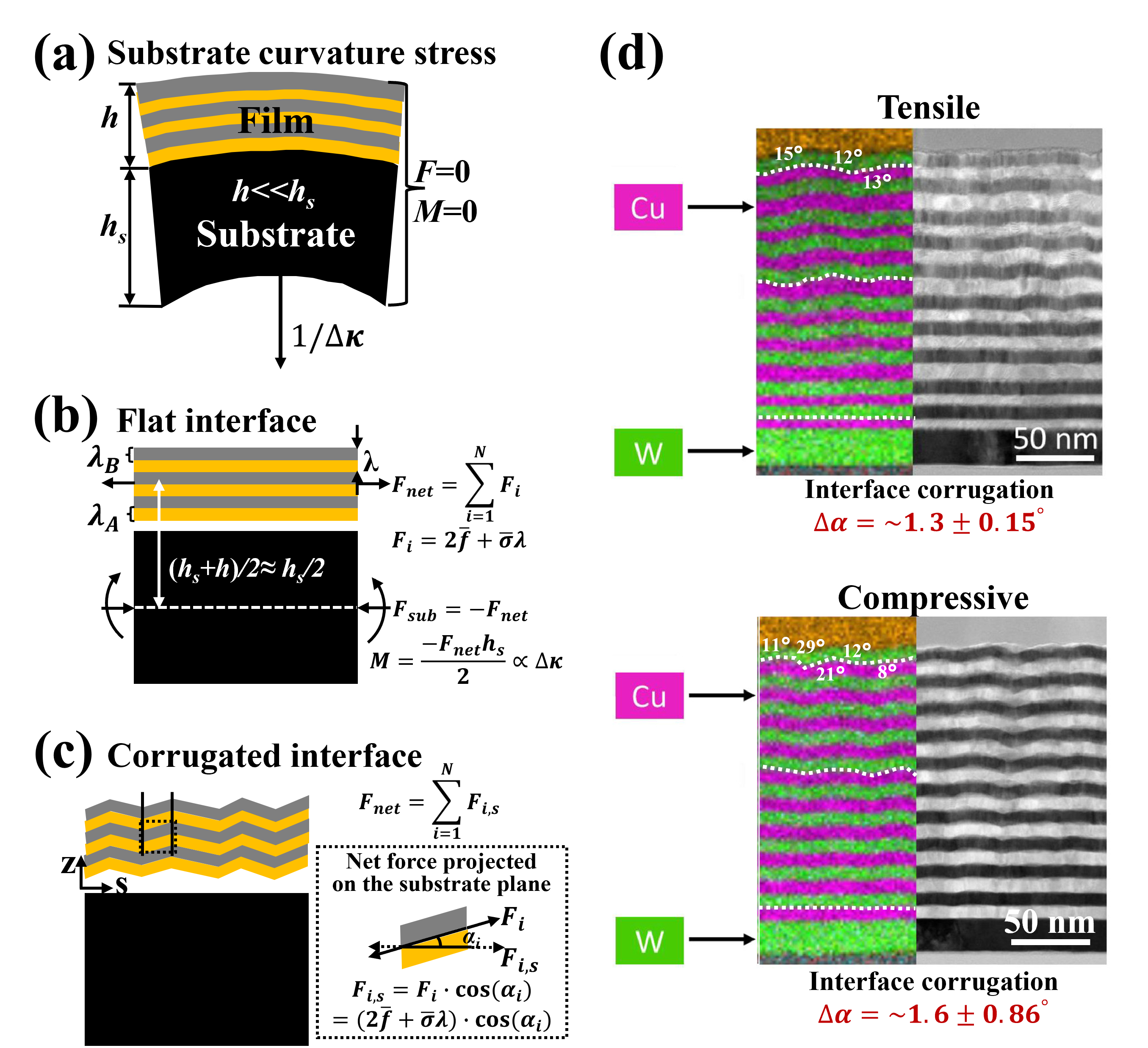}
\caption{\label{fig1} The schematic showing the (a) force and momentum balance in a curved NML/substrate system. Here $h$ denotes the film thickness, $h_{\textsubscript{s}}$ is the substrate thickness, $\lambda$ is the bilayer thickness, $\Delta \kappa$ represents the substrate curvature. (b) The force and momentum balance in a NML with flat interfaces, where $F_{\textsubscript{net}}$ is the net force per unit length due to the film’s intrinsic stresses and is counterbalanced by an equal and opposite force, $F_{\textsubscript{sub}}$ acting on the substrate. Within each bilayer, $F_{\textsubscript{i}}$ consists of two contributions: the interface stress, expressed as the average interface stress, $\overline{f}$ acting across the two interfaces, and the average residual stress $\overline{\sigma}$ in the Cu and W layers, multiplied by the bilayer thickness $\lambda$. (c) The force and momentum balance in a NML with corrugated interfaces, the angle $\alpha_{\textsubscript{i}}$ is defined between the highlighted portion of an interface (dotted box) and the substrate plane; $\alpha_{\textsubscript{i}}=0$ represents perfectly flat interfaces, whereas $\alpha_{\textsubscript{i}}=90$, corresponds to maximum interface corrugation. The tensile force responsible for bending the substrate is the component of the film force projected onto the substrate plane. (d) Energy-dispersive spectroscopy (EDS) image (left) and medium-angle annular dark-field scanning transmission electron microscopy (MAADF-STEM) image (right) showing cross-sectional views of a Cu/W NML under tension (top) and a Cu/W NML under compression (bottom), as reported in \cite{TRONCOSO2024115902}. Severe layer corrugations are observed near the film surface in both cases.}
\end{figure}

Two principal contributions to the intrinsic stress in NMLs are: (1) the residual stress caused by deviations of the lattice parameter from that of a strain-free bulk, and (2) the interface stress that arises from the high density of heterogeneous interfaces. For magnetron-sputtered NMLs, interface stress is commonly quantified by depositing NML coatings on flexible substrates. According to Ruud \textit{et al.} \cite{Ruud1993}, the film’s deposition induces curvature in the substrate, allowing one to measure the intrinsic thin-film (NML) stress, $\sigma^{\textsubscript{SC}}_{\textsubscript{NML}}$, as well as the net force (tensile or compressive) imposed on the substrate $F_{\textsubscript{net}}=\sigma^{\textsubscript{SC}}_{\textsubscript{NML}}h$, where $h$ is the total thickness of NML film composed of $N$ bilayers with thickness $\lambda$ each. As demonstrated in Fig. \ref{fig1}(b, c), this net force can also be expressed as the sum of forces contributed by two interfaces and two layers in a single bilayer: $F_\textsubscript{i}=2\overline{f}+\overline{\sigma}\lambda$, where $\overline{f}$ is the average interface stress and $\overline{\sigma}$ is the average biaxial residual stress in the bilayer. Consequently, one can derive the following equation from the force balance:
\begin{equation} 
\sigma^{\textsubscript{SC}}_{\textsubscript{NML}}h=\sum_{i=1}^N(2\overline{f}+\overline{\sigma}_0\lambda)=\left(\frac{2\overline{f}}{\lambda}+\overline{\sigma}\right)h.
\label{ruud}
\end{equation} 
The terms are defined as follows: $\sum_{i=1}^N\lambda=N\lambda=h$.

The interface stress can be then derived from:
\begin{equation} 
\overline{f} = (\sigma^{\textsubscript{SC}}_{\textsubscript{NML}}-\overline{\sigma})\lambda/2 
\end{equation}
where $\overline{f}=(f_{\textsubscript{AB}}+f_{\textsubscript{BA}})/2$, and $\overline{\sigma}$ can be approximated with the regular rule of mixtures, as A and B layers are positioned parallel to each other in the plane of stress. This results in the following equation:
\begin{equation} 
\overline{\sigma}=\frac{\lambda_{\textsubscript{A}}}{\lambda}\sigma_{\textsubscript{A}}^{\textsubscript{XRD}}+\frac{\lambda_{\textsubscript{B}}}{\lambda}\sigma_{\textsubscript{B}}^{\textsubscript{XRD}}.
\end{equation}
where $\lambda_\textsubscript{A}$ and $\lambda_\textsubscript{B}$ are thicknesses of individual layers forming the bilayer with a thickness of $\lambda=\lambda_\textsubscript{A}+\lambda_\textsubscript{B}$, and $\sigma_{\textsubscript{A}}^{\textsubscript{XRD}}$ and $\sigma_{\textsubscript{B}}^{\textsubscript{XRD}}$ are average residual stresses in two components A and B in NMLs commonly determined \textit{ex situ} by the X-ray diffraction (XRD) analysis.

Following this method, interface stresses have been measured in a variety of NML systems, including Ag/Ni \cite{Ruud1993,Schweitz1998,Birringer2009}, Ag/Fe \cite{Scanlon1995}, Ag/Cu \cite{Berger1995,Shull1996}, Au/Ni \cite{Schweitz2000,Labat2000}, and Cu/W \cite{DRUZHININ2021110002,Lorenzin2024}, with notably extreme values reported for Cu/W \cite{Lorenzin2024,hu2024uncovering}. Specifically, Cu/W NMLs deposited at an Ar pressure of 0.27 Pa exhibited interface stresses as high as 24 J/m$^{2}$, while those deposited at 2 Pa displayed interface stresses of -7 J/m$^{2}$. These NMLs also show increasingly pronounced layer corrugations from the substrate to the film surface \cite{MOSZNER2016345,DRUZHININ2021110002,lorenzin2022tensile}, which may introduce significant errors in interface stress measurements and potentially account for the extreme magnitudes observed in the Cu/W system. The origins of such layer corrugations in NMLs have been thoroughly explored, and corresponding mathematical models have been developed for different deposition regimes \cite{Huang2003}. However, \textit{no correction for layer corrugations in interface stress measurement equations has yet been proposed, nor has the associated measurement error been quantified; this constitutes the first objective of the present work.} Furthermore, interface roughness is a commonly reported phenomenon, often linked to the formation of kinks, ledges, or grain boundary grooves. Such features can affect the intrinsic stress distribution near the interface \cite{watkins2016neutron}. However, \textit{the effect of interface roughness on the measurement of interface stress has never been investigated, forming the second objective of this work.}

\section{Layer corrugation model}

In the presence of layer corrugations that do not alter cross-interface orientation relationships, the net force contributed by each layer and interface must be projected onto the substrate plane via the cosine of the corrugation angle $\alpha_{\textsubscript{i}}$, as shown in Fig. \ref{fig1}(c). In a simplified scenario where the corrugations are assumed to be periodic with constant amplitude and frequency within each bilayer $i$, the corresponding \textit{average} corrugation angle $\overline{\alpha}_\textsubscript{i}$ can be taken to depend only on the bilayer index, reflecting its distance from the substrate. Under these assumptions, the force balance becomes:

\begin{equation} 
\sigma^{\textsubscript{SC}}_{\textsubscript{NML}}h=\sum_{i=1}^N(2\overline{f}+\overline{\sigma}\lambda)\cdot \cos(\overline{\alpha}_\textsubscript{i}).
\label{corrected}
\end{equation} 

Fig. \ref{fig1}(d) presents cross-sectional images of a 10 nm/10 nm Cu/W NML deposited at both low and high Ar pressures \cite{lorenzin2022tensile}, illustrating that the layer corrugations are not uniform: layers are relatively flat near the substrate, becoming progressively more corrugated toward the surface. To capture this incremental change, one can write $\overline{\alpha}_\textsubscript{i}=\Delta\alpha\cdot i$, where $\Delta\alpha$ is the corrugation angle increment (assumed to be constant in zero-order approximation) and $i$ is the index of the bilayer across $z$ direction.
As explained in the Supplementary Materials, the summation in Eq. \ref{corrected} can be expressed as a function of $\Delta\alpha$ and $N$, so that

\begin{equation} 
\sigma_{\textsubscript{NML}}^{\textsubscript{SC}}h=(2\overline{f}+\overline{\sigma}\lambda) \frac{\sin(\frac{\Delta \alpha N}{2}) \cos(\frac{\Delta \alpha (N+1)}{2})}{\sin(\frac{\Delta \alpha}{2})},
\end{equation} 

Accounting that $h=N\lambda$, \textit{the final expression for interface stress in NMLs with layer corrugations is}:

\begin{equation} 
\overline{f}=\frac{\sigma^{\textsubscript{SC}}_{\textsubscript{NML}}N\lambda\sin(\frac{\Delta \alpha}{2})}{2\sin(\frac{\Delta \alpha N}{2}) \cos(\frac{\Delta \alpha (N+1)}{2})}-\frac{\overline{\sigma}\lambda}{2}.
\end{equation} 

The average corrugation angles at the top bilayers of the two experimental samples shown in Fig. \ref{fig1}(d) were estimated to be 16$^{\circ}$ for the NML under compression and 13$^{\circ}$ for the NML under tension. Dividing these average angles by the number of bilayers (N=10) then provides the incremental corrugation angles for each sample i.e., 1.6$^{\circ}$ and 1.3$^{\circ}$, respectively. However, these values should be regarded as approximate, since they rely on visual inspection of EDS images and thus carry substantial measurement uncertainty. Consequently, we cannot definitively conclude that a sample deposited under compression exhibits more pronounced layer corrugations than the one deposited under tension. To increase the accuracy of the corrugation angle increment derivation, the combination of cross-section imaging with profilometry and/or atomic force microscopy of the top nanolayer is preferable. In the present study, the order of the corrugation angle magnitude was of the main interest, so only the cross-section investigation was used with a limited set of sampled corrugation angles. 

\section{Interface roughness model}

\begin{figure}[H]
\centering
\includegraphics[width=0.75\textwidth,clip,trim=0cm 0cm 0cm 0cm]{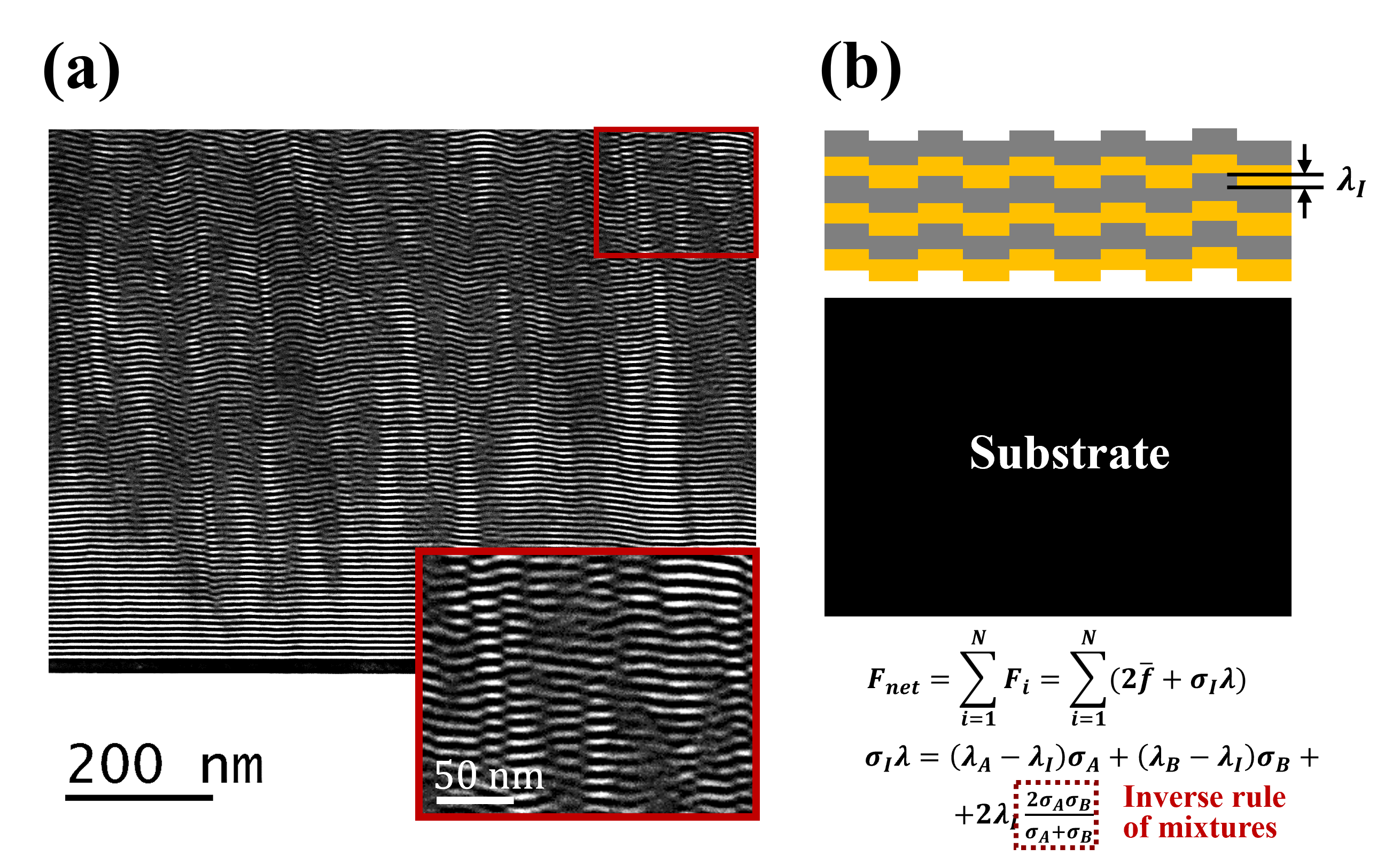}
\caption{\label{fig2} (a) Cross-sectional STEM image of a Cu/W NML grown at 0.56 Pa Ar pressure; each layer is 5 nm and the number of repetitions was 100 for a total thickness of 1 $\text{$\mu$}$m. More details on sample preparation are given in Ref. \cite{MOSZNER2016345}.
Thin lamellas of the NML in the as-deposited were cut using a HITACHI NB5000 FIB-SEM system. Their further imaging was done as a part of this work using a Hitachi-HD2700 Scanning Transmission Electron Microscope (STEM), equipped with a Cs corrector operating at 200 kV.
(b) The schematic showing the interface roughness within NMLs deposited on a substrate, where $\lambda_{\textsubscript{I}}$ is the average interface roughness.}
\end{figure}

Layer corrugations typically span multiple grains and are readily visible in SEM/EDS images (see Fig. \ref{fig1}(d) for example). By contrast, interface roughness—arising from kinks, ledges, and grain boundary grooves—tends to exhibit periods on the order of the average grain size. Although the amplitude of interface roughness is generally much smaller than that of layer corrugations, it can still become pronounced in certain cases, particularly when columnar growth emerges \cite{Brandt2021}. Such microstructure commonly emerges in NMLs with small layer thicknesses and a large number of repetitions. As shown in Fig. \ref{fig2}(a) for 5 nm/5 nm Cu/W NMLs, the effective facet shear along the growth direction could reduce extreme layer corrugations and align individual facets back to be parallel to the substrate plane, however disrupting the continuity of individual layers and inducing substantial interface roughness. However, such extreme cases are not practical for any interface stress measurements as they result in rather an unpredictable interplay between layer corrugations and interface roughness. 

In the presence of only interface roughness, the cross-interface orientation relationships generally remain unchanged so that the average interface stress can be treated as equivalent to that of flat interfaces. It is important to note, however, that interface roughness is conceptually distinct from interfacial intermixing, which operates at the atomic scale and can strongly influence interface stress \cite{hu2024uncovering}. Instead, interface roughness primarily alters the intensity and broadening of XRD peaks, while leaving the peak position—critical for determining the residual stress via XRD largely unaffected. Although the interface roughness itself does not directly modify the XRD-based stress measurements in each individual layer (e.g., Cu and W), it can affect how the bilayer residual stresses are averaged to yield the total multilayer (NML) stress. As shown in Fig. \ref{fig2}(b), the average stress along the substrate plane for individual bilayers with rough interfaces can be computed using a direct rule of mixtures:

\begin{equation}
\overline{\sigma}=\frac{(\lambda_{\textsubscript{A}}-\lambda_{\textsubscript{I}})}{\lambda}\sigma_{\textsubscript{A}}^{\textsubscript{XRD}}+\frac{(\lambda_{\textsubscript{B}}-\lambda_{\textsubscript{I}})}{\lambda}\sigma_{\textsubscript{B}}^{\textsubscript{XRD}}+2\frac{\lambda_{\textsubscript{I}}}{\lambda}\sigma_{\textsubscript{I}},
\end{equation}
where $\lambda_{\textsubscript{I}}$ is the average thickness of a single interface region and $\sigma_{\textsubscript{I}}$ is the average stress in the interface region. The latter is composed of alternating phases along the substrate plane, which demands the use of the inverse rule of mixtures:
\begin{equation}
\sigma_{\textsubscript{I}}=\frac{2\sigma_{\textsubscript{A}}^{\textsubscript{XRD}}\sigma_{\textsubscript{B}}^{\textsubscript{XRD}}}{\sigma_{\textsubscript{A}}^{\textsubscript{XRD}}+\sigma_{\textsubscript{B}}^{\textsubscript{XRD}}}.
\end{equation}
It is reasonable to assume that the average interface roughness is nearly constant from interface to interface. The average roughness per interface (or interface disorder) can be estimated by the modulated XRD intensity oscillations, i.e. satellite peaks, typically observed for a periodic nanolayered structure along the $z$-axis \cite{MOSZNER2016345, lorenzin2022tensile, Ariosa, CANCELLIERI2024114362}. As discussed in \cite{lorenzin2022tensile}, the parameter $\mu$ derived from the model fit of the XRD data, contains the information on the interface disorder quantified as the number of lattice planes displacement along the $z$-axis, with respect to the plane of the nominal thickness. The average interface roughness can be calculated as $\sqrt{\mu} \cdot d$ where $d$ is the average distance of the planes along the growth direction (e.g. calculated as ($d_{\text{Cu}(111)}$+$d_{\text{W}(110)}$)/2). 
\begin{table}[H]
\centering
\caption{Parameters from the XRD model fit from ref. \cite{lorenzin2022tensile} used to calculate the interface roughness.}
\label{muTable}
\begin{tabular}    
{|c@{\hspace{0.2cm}}|c@{\hspace{0.18cm}}|c@{\hspace{0.18cm}}|c@{\hspace{0.18cm}}|}
\hline
Sample & $\mu$ & $d$ (nm) & Roughness $\lambda_{\textsubscript{I}}$(nm)     \\
\hline
Tensile & 121 & 0.213 \AA & 2.3  \\
\hline
Compressive & 6 & 0.217 \AA & 0.5    \\
\hline
\end{tabular}
\end{table}

For the two experimental samples depicted in Fig. \ref{fig1}(d), parameters $\mu$ and $d$ are listed in Table \ref{muTable}. The NML under tension exhibits a larger estimated roughness of approximately 2.3 nm, whereas the NML under compression shows a comparatively smaller value of about 0.5 nm. This quantitative result is also consistent with STEM images provided in Fig. \ref{fig1}(d), where the high interface roughness in the tensile sample leads to the perceived variation in layer thicknesses while in reality the deposition time for both Cu and W was carefully calibrated to deliver the same layer thicknesses across NMLs.  

\begin{table}[H]
\centering
\caption{Interface stress values derived with different models for Cu/W NMLs. Residual and film stresses are taken from \cite{lorenzin2024effect}, layer thicknesses $\lambda_{\textsubscript{Cu}}=\lambda_{\textsubscript{W}}=\lambda/2=10$ nm and number of bilayers $N=10$ for both compressive and tensile samples.}
\label{discuss}
\begin{tabular}{|c@{\hspace{0.5cm}}|c@{\hspace{0.3cm}}|c@{\hspace{0.3cm}}|}
\hline
Sample  & Compressive & Tensile \\
\hline
Ar pressure (Pa) & 0.27 & 2 \\
Cu XRD stress $\sigma_{\textsubscript{Cu}}^{\textsubscript{XRD}}$ (GPa) & -1.80 & 0.66 \\
W XRD stress $\sigma_{\textsubscript{W}}^{\textsubscript{XRD}}$ (GPa) & -3.84 & 1.50 \\
Film stress $\sigma^{\textsubscript{SC}}_{\textsubscript{NML}}$ (GPa) & -0.441 & 0.403 \\
Corrugation increment $\Delta\alpha$ (\textdegree) & 1.6 & 1.3\\
Interface roughness $\lambda_\textsubscript{I}$ (nm) & 0.5 & 2.3 \\
\hline
\multicolumn{3}{|c|}{Interface stress (J/m$^2$)} \\
\hline
Ruud et al. model, Eqs. (2),(3) & 23.79 & -6.77 \\
Corrugation model, Eqs. (6),(3) & 23.72 & -6.73 \\
Roughness model, Eqs. (2),(7),(8) & 23.61 & -6.40 \\
Combined model, Eqs. (5),(7),(8) & 23.54 & -6.36 \\
\hline
\end{tabular}
\end{table}

\section{Revised interface stresses}

The interface stresses for the two 10 nm/10 nm Cu/W NMLs shown in Fig. \ref{fig1}(d) were estimated using several models: the standard approach by Ruud \textit{et al.}, our layer corrugation model, our interface roughness model, and finally a combined model that accounts for both effects. The results, summarized in Table \ref{discuss}, confirm that neglecting corrugation and roughness tends to overestimate the interface stress. However, the magnitude of this overestimation is small, on the order of 0.4 J/m$^{2}$, which corresponds to a relative error of about 6\% compared to the standard model. This small discrepancy lies well within the experimental uncertainty propagated from both XRD residual stress measurements and the substrate-curvature-based film stress measurements \cite{Lorenzin2024}. This highlights the reliability of the previous experimental measurements and indicates that \textit{the extreme magnitudes of interface stress reported for Cu/W NMLs are not attributed to layer corrugation and interface roughness}. 

\begin{figure}[H]
\centering
\includegraphics[width=0.75\textwidth,clip,trim=0cm 4.2cm 0cm 0cm]{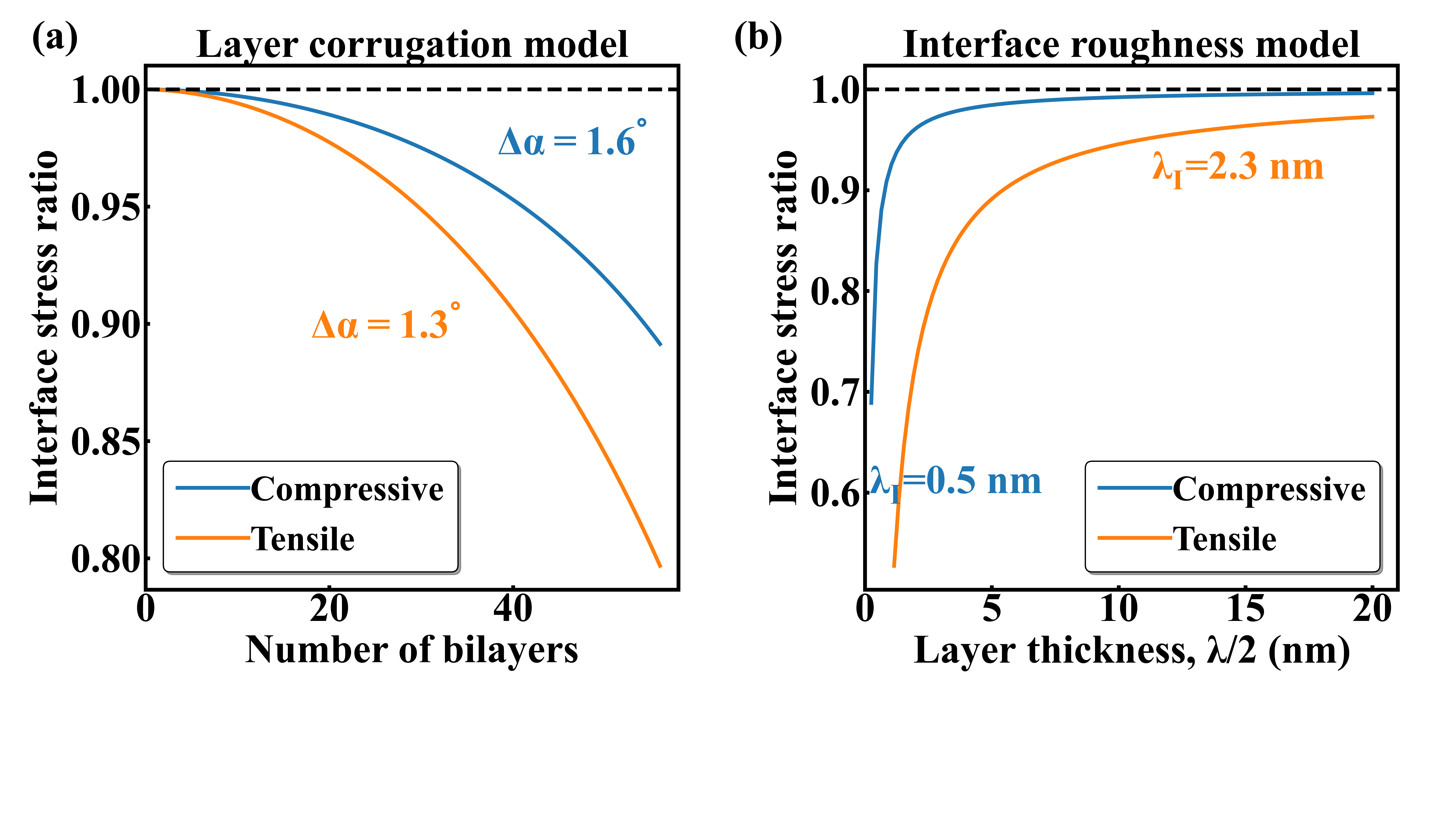}
\caption{\label{fig3} The variation of the interface stress ratio (a) with the number of bilayers, and (b) with layer thicknesses. In (a), the ratio was computed using the layer corrugation model and the Ruud \textit{et al.}'s model (reference horizontal dashed line at value 1), while in (b), the ratio was computed using the interface roughness model and the Ruud \textit{et al.}'s model.}
\end{figure}

Despite the modest magnitude of corrugation and roughness corrections in our particular samples, it remains crucial to establish best practices for accurate interface stress measurements in NMLs, given the high density of interfaces and the significant role that interface stress can play in defining mechanical, thermal, and other functional properties. From our layer corrugation model (Eq. (6)), one sees that the correction to the measured film stress depends on the number of bilayers, $N$. In the original formulation of Ruud \textit{et al.}, $N$ must be sufficiently large so that the top free surface and the NML/substrate interface contributions become negligible. However, our results in Fig. \ref{fig3}(a) demonstrate that, as $N$ increases, the error in extracted interface stress also grows if corrugation is not accounted for. Balancing these two trends suggests an optimal $N$ of around 10 for bilayers as it ensures that the corrugation effect is negligible while the number of interfaces within NML is one order of magnitude bigger than other interfaces (the NML/substrate interface and the free surface). 

In addition, the interface roughness model (Eq. (7)) reveals that its correction to the average residual stress in the bilayers depends strongly on individual layer thicknesses. NMLs should be prepared with each layer being thick enough to form a continuous film rather than disconnected islands or grains, but still thin enough that the overall NML thickness remains small relative to the substrate thickness so that the requirement introduced in Fig. 1(b) is obeyed. However, as we can see from the interface roughness model, such a requirement should be further extended to reflect that the layer thickness should be substantially larger than the interface roughness to ensure an accurate measurement of interface stress. As there are no specific advantages of having different thicknesses of A and B layers, we focus on the case of equal thicknesses of A and B layers. Fig. \ref{fig3}(b) demonstrates that the corrected interface stress diverges more significantly from the Ruud \textit{et al.} model prediction as the layer thickness approaches the characteristic interface roughness scale. As the interface roughness seems to be sensitive to deposition conditions (see Table 1), it is hard to estimate the optimal layer thickness so that the effect of interface roughness can be completely neglected. Yet in practice, one can conveniently estimate interface roughness by analyzing XRD line profiles; we thus recommend employing Eq. (7) for a straightforward roughness-based correction for a given NML system.

It is also important to note that layer corrugation and interface roughness are not entirely independent structural features. For instance, the layer corrugation does not simply increase indefinitely as the layer number grows; at some point, large corrugation amplitudes can be relaxed by severe interface roughness, as illustrated in Fig. 2(a). In this sense, the interplay between these two factors can modulate the final interface morphology and thus the effective stress state in the film.

Altogether, these findings underscore that while both layer corrugation and interface roughness can be systematically accounted for in determining the interface stress, their effects on Cu/W NMLs are relatively small compared to experimental uncertainties. In other words, these meso- to microscale geometric features do not explain the extremely large interface stresses reported experimentally. This calls attention to the importance of refining not only the experimental procedures for measuring interface stresses (so that various geometric artifacts are minimized) but also the interpretation of interface stress in terms of the atomic-scale features that ultimately dominate.

\begin{figure}[H]
\centering
\includegraphics[width=0.55\textwidth,clip,trim=0cm 0cm 0cm 0cm]{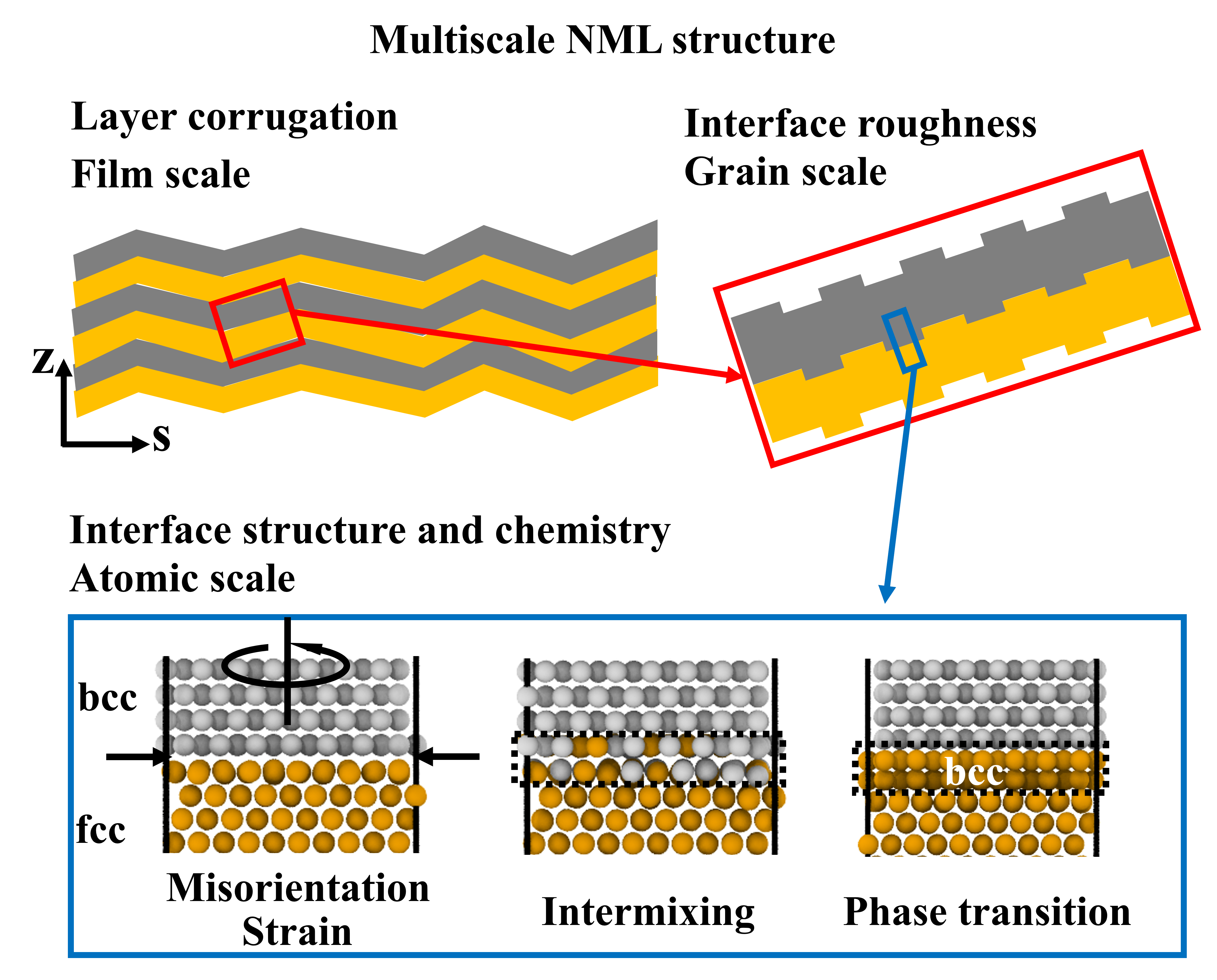}
\caption{\label{fig4} The schematic showing the multiscale structure of NMLs.}
\end{figure}

Indeed, interface stress is fundamentally defined at the atomic scale, and NMLs exhibit a multiscale hierarchy of structural features, ranging from layer corrugation and interface roughness on the nanoscale to in-plane orientation, in-plane residual strains (multilayer and mismatch strains), local chemical intermixing, and possible metastable phase formation within just a few atomic layers of the interface (see Fig. \ref{fig4}). These latter phenomena can significantly surpass the geometric corrections in magnitude. For instance, in the classical immiscible Ag/Ni system, textured interfaces exhibit negative interface stress \cite{Ruud1993,Schweitz1998}, whereas untextured interfaces show positive interface stress \cite{Birringer2009}. While such transition can be attributed to interface misorientation, it can also be the result of interface intermixing \cite{hu2024uncovering,Clemens2000}. Nevertheless, it is nearly impossible to establish the correspondence between interface stress and its state using experimental techniques, thus making the interpretation of experimental measurements very challenging. Contrary, the emergence of machine learning interatomic potentials in the field of atomistic simulations recently enabled fast screening and characterization of different interface misorientations and states with \textit{ab initio} accuracy, revolutionizing the interpretation of experimentally-measured interface stresses \cite{hu2024uncovering}.   

Particularly for Cu/W NMLs in which the extreme magnitudes of interface stresses have been observed \cite{lorenzin2024effect}, Hu \textit{et al.} conducted molecular dynamics simulations on \{111\}$_{\text{Cu}}$/\{110\}$_{\text{W}}$ interfaces with different in-plane orientations and interfacial chemistry, and observed a near 30\% difference by comparing the measured interface stress for a sharp, unstrained interface of random in-plane orientation with a Nishiyama-Wasserman (NW) and a Kurdjumov-Sachs (KS) interface of the same condition. Moreover, small in-plane residual strains (on the order of $\pm$1\%) can shift the measured interface stress by around 24\% for the three interfaces mentioned above \cite{hu2024uncovering}. However, in-plane orientations and residual strains are still insufficient to explain the over 400\% increase in the measured interface stresses from a sharp interface. By contrast, intermixing and phase transitions at the interface can reach stress magnitudes comparable to experimental observations in Cu/W NMLs \cite{hu2024uncovering}. As one can see, \textit{such a powerful atomic-level analysis provides a detailed quantitative interpretation of interface stress values, signifying the importance of accurate experimental measurement of interface stress (now more than ever), accounting for geometric considerations such as layer corrugations and interface roughness investigated in this work}. 

\section{Conclusions}

In summary, we refine the measurement of interface stress in nanomultilayers (NMLs) by introducing new phenomenological models that explicitly account for layer corrugation and interface roughness. By analyzing available experimental samples, we show that neglecting these mesoscale features can introduce sizable errors up to 0.4 J/m$^2$ in interface stress calculations. Nevertheless, the extreme stress magnitudes measured previously in Cu/W NMLs are more plausibly explained by atomic-scale factors such as intermixing and metastable phase formation than by microstructural features such as layer corrugations and surface roughness. Nevertheless, the latter can induce substantial errors in interface stress measurement if the number of deposited bilayers is too large (much more than 10) or the layer thickness is near the same order of magnitude as interface roughness. In the latter case, we recommend direct estimation of interface roughness from XRD profiles and calculating interface stress with the updated model from this work. Overall, our findings highlight the multiscale nature of stress development in NMLs and underscore the need to address structural features at different length scales for a more complete understanding of intrinsic stresses within NMLs. 

\section{Acknowledgments}
This research was supported by NCCR MARVEL, a National Centre of Competence in Research, funded by the Swiss National Science Foundation (grant number 205602). We thank Dr. Giacomo Lorenzin for sample preparation and for providing raw data. We acknowledge Dr. Mariusz Andrzejczuk for the STEM measurements performed at Warsaw University of Technology. We also thank Lars Jeurgens for fruitful discussions.

\appendix

\section{Appendix A: Derivation of the sum \( S = \sum_{k=1}^{n} \cos(a k) \)}
To find the sum \( S = \sum_{k=1}^{n} \cos(a k) \) step by step, we'll use trigonometric identities and complex numbers. Here's how to derive the formula:
\

**Step 1: Express the Cosine Function Using Complex Exponentials**

We can express the cosine function in terms of complex exponentials:

\[
\cos(a k) = \text{Re}\left\{ e^{i a k} \right\}
\]

So, the sum becomes:

\[
S = \sum_{k=1}^{n} \cos(a k) = \text{Re}\left\{ \sum_{k=1}^{n} e^{i a k} \right\}
\]

**Step 2: Compute the Geometric Series**

The sum inside the real part is a geometric series:

\[
\sum_{k=1}^{n} e^{i a k} = e^{i a}\sum_{k=0}^{n-1} e^{i a k}= e^{i a} \left( \frac{1 - e^{i a n}}{1 - e^{i a}} \right)
\]

**Step 3: Separate Numerator and Denominator**

Let’s denote:

- Numerator: \( N = e^{i a} - e^{i a(n+1)} \)

- Denominator: \( D = 1 - e^{i a} \)

\

**Step 4: Express Numerator and Denominator in Trigonometric Form**

Using Euler's formula \( e^{i \theta} = \cos \theta + i \sin \theta \), we have:

\( N = [\cos a - \cos((n+1)a)] + i[\sin a - \sin((n+1)a)] \)

\( D = (1 - \cos a) - i \sin a \)

\

**Step 5: Apply Trigonometric Identities**

We use the following identities:

1. \( \cos A - \cos B = -2 \sin\left( \frac{A+B}{2} \right) \sin\left( \frac{A-B}{2} \right) \)

2. \( 1 - \cos A = 2 \sin^2\left( \frac{A}{2} \right) \)

3. \( \sin A - \sin B = 2 \cos\left( \frac{A+B}{2} \right) \sin\left( \frac{A-B}{2} \right) \)

Applying these:

- \( N_{\text{real}} = -2 \sin\left( \frac{(n+2)a}{2} \right) \sin\left( \frac{n a}{2} \right) \)

- \( N_{\text{imag}} = 2 \cos\left( \frac{(n+2)a}{2} \right) \sin\left( \frac{n a}{2} \right) \)

- \( D_{\text{real}} = 2 \sin^2\left( \frac{a}{2} \right) \)

- \( D_{\text{imag}} = - \sin a \)
\\

**Step 6: Multiply the numerator and denominator by $D_{\text{real}}-i D_{\text{imag}}$**

The magnitude squared of the denominator is:
\[
|D|^2 = D_{\text{real}}^2 + D_{\text{imag}}^2 = 4 \sin^2\left( \frac{a}{2} \right)
\]
The numerator is:
\\

$\left(N_{\text{real}}+i N_{\text{imag}}\right) \left(D_{\text{real}}-i D_{\text{imag}} \right)=N_{\text{real}}D_{\text{real}}+N_{\text{imag}}D_{\text{imag}}+i \left(N_{\text{imag}}D_{\text{real}}-N_{\text{real}}D_{\text{imag}} \right)$
\\

**Step 7: Compute the Real Part of the Sum**

The real part of the sum is:

\[
S = \frac{N_{\text{real}} D_{\text{real}} + N_{\text{imag}} D_{\text{imag}}}{|D|^2}
\]

Substituting the expressions from Step 5:

\[
S = \frac{4 \sin\left( \frac{n a}{2} \right) \sin\left( \frac{a}{2} \right) \cos\left( \frac{(n+1) a}{2} \right)}{4 \sin^2\left( \frac{a}{2} \right)}
\]

Simplify:

\[
S = \frac{\sin\left( \frac{n a}{2} \right) \cos\left( \frac{(n+1) a}{2} \right)}{\sin\left( \frac{a}{2} \right)}
\]

**Step 8: Final Result**

The sum of the series is:

\[
\sum_{k=1}^{n} \cos(a k) = \frac{\sin\left( \dfrac{n a}{2} \right) \cos\left( \dfrac{(n+1) a}{2} \right)}{\sin\left( \dfrac{a}{2} \right)}
\]

\bibliography{main}
\end{document}